\documentclass[]{acmart}
\AtBeginDocument{%
  \providecommand\BibTeX{{%
    \normalfont B\kern-0.5em{\scshape i\kern-0.25em b}\kern-0.8em\TeX}}}

\setcopyright{rightsretained}
\copyrightyear{2024}
\acmYear{2024}
\acmDOI{XXXXXXX.XXXXXXX}

\acmConference[FAccT '24]{June 03--06, 2024}{Rio de Janeiro, Brazil}
\acmSubmissionID{7}

\begin{document}

\title{Data Feminism for AI}

\author{Lauren Klein}
\authornote{Both authors contributed equally to this research.}
\email{lauren.klein@emory.edu}
\orcid{0002-1511-0910}
\affiliation{%
  \institution{Emory University}
  \city{Atlanta}
  \state{GA}
  \country{USA}
}

\author{Catherine D'Ignazio}
\authornotemark[1]
\email{dignazio@mit.edu}
\orcid{}
\affiliation{%
  \institution{MIT}
  \city{Cambridge}
  \state{MA}
  \country{USA}
}

\renewcommand{\shortauthors}{Klein and D'Ignazio}

\begin{abstract}
This paper presents a set of intersectional feminist principles for conducting 
equitable, ethical, and sustainable AI research. In \textit{Data Feminism} (2020), we offered seven principles for examining and challenging unequal power in data science. Here, we present a rationale for why feminism remains deeply relevant for AI research, rearticulate the original principles of data feminism with respect to AI, and introduce two potential new principles related to environmental impact and consent. Together, these principles help to 1) account for the unequal, undemocratic, extractive, and exclusionary forces at work in AI research, development, and deployment; 2) identify and mitigate predictable harms in advance of unsafe, discriminatory, or otherwise oppressive systems being released into the world; and 3) inspire creative, joyful, and collective ways to work towards a more equitable, sustainable world in which all of us can thrive.
\end{abstract}





\received[submitted]{22 January 2024}
\received[accepted]{30 March 2024}

\maketitle

\section{Introduction}

\textit{Data Feminism} \cite{D’Ignazio_Klein_2020} was published in March 2020, in the first week of what would become a world-altering pandemic, and in the wake of over a decade of increasing awareness of the power of data when collected, analyzed, and deployed. Our motivation for writing the book was the overabundance of evidence of the power of data, and of how that power was being wielded unequally. More specifically, it was being wielded by corporations, governments, and other well-resourced institutions to enhance their own power and profit, with significant personal, political, and financial costs for everyone else. There was already a rich scholarly conversation about how data was being used to amplify existing structural inequalities \cite{Broussard_2018, O’Neil_2016, Eubanks_2018, Benjamin_2019, Buolamwini_Gebru_2018, Walter_2013}. The contribution of \textit{Data Feminism} was to show how feminism, because of its analytic focus on the root causes of structural inequalities, could help challenge and rebalance that power. The seven principles of data feminism–examine power, challenge power, rethink binaries and hierarchies, elevate emotion and embodiment, consider context, embrace pluralism, and make labor visible–were intended to operationalize what we saw as the most relevant tenets of feminist thinking for data science. Our goal was to provide a clear set of guidelines and examples for people working with data, who wanted to work with data, or who wanted to refuse to work with data on political or personal grounds. We wanted to show how feminism was not only relevant but essential to data science and, to model how data scientists, computer scientists, digital humanists, policymakers, urban planners, journalists, educators, students, and others could put feminism into practice in their work.

Since the publication of the book, the principles of \textit{Data Feminism} have been taken up across academia \cite{Williams_2021, Byrne_2022, Darian_Chauhan_2023, Hatch_Raymond_Teresa_Howell_2023,  Lee_Pimentel_Bhargava_D’Ignazio_2022, Lean_2021, Turculet_2023} and in the public sector  \cite{Data_Genero_2023, Neema_Chenai_Garnett_2023, Kashyap_Singh_2021, Data_Feminism_2022, D’Ignazio_2022, Klein_Marshall_2022}. But as the conversation has shifted from data science to AI, we see a need to revisit these principles. This paper presents a rationale for why feminism remains deeply relevant for AI research, rearticulates the original principles of data feminism with respect to AI, and introduces the possibility of two additional principles, related to environmental impact and consent, in order to help to 1) account for the unequal, undemocratic, extractive, and exclusionary forces at work in AI research, development, and deployment; 2) identify and mitigate predictable harms in advance of unsafe, discriminatory, or otherwise oppressive systems being released into the world; and 3) inspire creative, joyful, and collective ways to work towards a more equitable, sustainable world in which all of us can thrive..

\section{Background}

\subsection{What is Feminism?}
Feminism has a long, varied, and often contested history. While it exceeds the scope of this paper to summarize the entirety of this history, it is important to clarify the definition of feminism we mobilize here. At its most basic level, feminism entails a belief in the equality of all genders. This includes women and men, as well as Two Spirit, genderqueer, travesti, nonbinary people, and more. But until gender equality is realized in the world, our feminism also requires organized activity to make this goal of equality a reality. A third aspect of our feminism derives from its intellectual heritage, and a crucial part of this heritage is \textit{intersectional feminism}, which comes to us from the work of women of color feminists, and Black feminists in the United States in particular. The contributions of intersectional feminism are twofold: first, to bring additional facets of social difference to the conversation about gender inequality, including but not limited to racial and economic inequality; and second to insist that we concern ourselves with structural power: the reasons why people experience privilege on the one hand, or oppression on the other. \textit{Intersectional feminism} offers models (in the conceptual sense, not the machine learning sense) that explain the causal mechanisms of complex systems of power and guide action to transform them towards justice. These include the Combahee River Collective’s observation about “interlocking systems of oppression” \cite{Combahee}, Patricia Hill Collins’s formulation of the “matrix of domination” \cite{Hill_Collins_2000}, and Kimberlé Crenshaw’s term “intersectionality” \cite{Crenshaw}. Note that these intersectional feminist models of power are not the only models of structural power that exist; others have theorized the workings of structural inequality from the perspective of capitalism, colonialism, and so on. We are drawn to intersectional feminist theories of power because of how they bridge personal experience and structural frameworks, and because they are explicit about their goal: understanding present imbalances of power in the world so that they can be challenged, rebalanced, and changed.  

\subsection{Feminism Today}

In the years since \textit{Data Feminism} was published, there have been several significant alterations to the social and political fabric of the United States and the world, many of which bring feminist considerations to the fore. Most directly, in the US, where the authors of this paper are located, we have experienced the overturning of the constitutional right to abortion, which has set off a cavalcade of increasing incursions into the autonomy and privacy of those with child-bearing bodies. These long-standing feminist concerns now play out in digital (if not exclusively AI-driven) spaces, as personal data has become a key legal weapon wielded against people seeking abortions \cite{Sandvik_2023}. Cases such as these also underscore the close relation between reproductive justice and trans justice, in that attacks on the bodily autonomy of some are attacks on the autonomy of all. It is not a coincidence that restrictions on abortion access have been accompanied by restrictions on gender-affirming care, and other legal efforts to police the bodies of trans people. %In 2023 alone, the ACLU tracked 510 anti-LGBTQ+ bills across the US. More than a quarter of these were focused on denying trans people preventative health care \cite{Mapping_2023}. 
A key feminist lesson here, born from over a century of exclusionary history, is that these are interconnected struggles, and we cannot defend bodily autonomy with an essentialist definition of “women” %White feminists, in particular, have a well-documented history of compromising their stated values when their own rights and privileges were on the line 
\cite{Schuller_2021}. 
%We must ensure that any defense of reproductive rights does not further entrench exclusionary definitions of womanhood or of gender. 
A related lesson comes to the US from Latin America. There, the decades-long “marea verde” (green wave) of feminist activism has been successful in large part due to its expansiveness and economic populism as well as its insistence on linking labor issues to gender issues \cite{Revilla_Blanco_2019}. As the US rolled back abortion rights, for example, Argentina legalized abortion and it was decriminalized in Colombia and Mexico, some of the most populous countries in the hemisphere. These successes were the result of several decades of protest and activism. 
%Feminist and LGBTQ+ movements in Argentina have also achieved rights to change one's gender marker, the inclusion of a third gender marker, a trans/travesti employment quota in the federal government and more.  Feminists in the US have much to learn from our compañeras and compañeres from the South.   

Attacks on women and trans people are not happening in a vacuum. Communications scholars W. Lance Bennett and Marianne Kneuer recently argued that we are witnessing the intensification of "illiberal public spheres" around the world \cite{Bennett_Kneuer_2023}. These are characterized by a set of anti-democratic communications tactics that include the denigration and exclusion of minoritized people, targeted attacks on the press and political institutions, and transgression of norms of civil discourse. Such tactics are enabled and amplified by the social media platforms owned by Big Tech, whose business model elevates attention above all other metrics, such that they profit richly off extremism, threats, spectacle, and lies. The success of this corporate partnership with right-wing extremism is evident in the proliferation of book bans \cite{Magnusson}, the demonization of DEI and critical race theory \cite{Crenshaw_2022}, and mob tactics to doxx, intimidate, misrepresent, and otherwise silence people that study misinformation, speak out against racial and gender violence, or rule against insurrectionists \cite{Edmonds_2024}. 
%What does it say about the condition of our democracy if the NAACP, one of the oldest civil rights organizations in the US, has to issue a travel ban for the safety of Black people? They state, "Florida is openly hostile toward African Americans, people of color and LGBTQ+ individuals. Before traveling to Florida, please understand that the state of Florida devalues and marginalizes the contributions of, and the challenges faced by African Americans and other communities of color." 
At the same time, we have seen the continued work of feminists, including Latin American, Indigenous, and abolitionist feminists in the US and globally, doing the work of imagining alternate worlds \cite{Imagining_Abolition_2021, Davis_2022, Ricaurte2019, Simpson_2017}. 
%This paper represents our own first attempt to imagine a more equitable, sustainable world for AI research in which all of us can thrive.
This paper represents our own first attempt to imagine a more equitable, sustainable, feminist world for AI research.

%%These include direct abolitionist-feminist interventions into carceral systems \cite{Imagining_Abolition_2021, Davis_2022}, as well as broader reimaginings of society \cite{Benjamin_2022, Benjamin_2024}. We have also seen abolitionist-feminist ideas enter into pedagogy \cite{Earl_2021} and policy around data and AI \cite{Databite_No_129}. We hold the idea of “abolition as method” in mind as we move forward \cite{gabriel_2022}

\section{Related Work}

\subsection{Feminism and FAccT}

Within the FAccT community, we have already seen examples of how feminism can play a substantive role in guiding AI/ML research. As early as 2021, Leila Marie Hampton \cite{Hampton_2021} introduced Black feminism as a lens through which to understand and critique algorithmic oppression. That same year, Hancox-Li and Kumar \cite{Hancox-Li_Kumar_2021} introduced the concept of feminist epistemology to the FAccT community, offering suggestions on how to align AI/ML research with feminist ideas about the value of situated knowledge \cite{Haraway_1988} and of multiple ways of knowing more broadly. In subsequent years, work at FAccT has demonstrated how a suite of feminist concepts drawn from Black feminism, feminist STS, and feminist new materialism could be applied to ML research \cite{Klumbytė_Draude_Taylor_2022} and how intersectionality had thus far been (weakly) operationalized in AI fairness research \cite{Kong_2022}. 

Feminism has also informed a range of applied work, including research on the harms of online advertising systems \cite{Sampson_Encarnacion_Metaxa_2023} and workplace surveillance \cite{Chowdhary_Kawakami_Gray_Suh_Olteanu_Saha_2023}, as well as participatory processes for ML design \cite{Suresh_Movva_Dogan_Bhargava_Cruxen_Cuba_Taurino_So_D’Ignazio_2022}. Other papers at FAccT have explored the topic of gender more concretely, including the issue of gender bias in public-facing tools (e.g. autocomplete \cite{Leidinger_Rogers_2023}) and research methods (e.g. NLP \cite{Devinney_Björklund_Björklund_2022}), as well as in the field of computer science itself \cite{Cheong_Leins_Coghlan_2021}. But this work remains a small minority. In their 2022 meta-analysis of AI ethics research conducted at FAccT, Birhane et al. concluded that “the field would benefit from an increased focus on ethical analysis grounded in concrete use-cases, people’s experiences, and applications as well as from approaches that are sensitive to structural and historical power asymmetries” \cite{Birhane_Ruane_2022}. The principles offered here respond to that call by providing a feminist framework to structure this necessary work. 

\subsection{Feminism and AI}

Looking more broadly, scholarship on feminism and AI stretches back at least to Alison Adam's 1998 book, \textit{Artificial Knowing} \cite{Adam1998-hu}, which, as Keyes and Creel \cite{Keyes_Creel_2022} remind us, used feminist epistemology to challenge the presumed universality of AI research of the time. One of the most widely-cited scholarly papers in AI ethics today, by Joy Buolamwini and Timnit Gebru, employed an intersectional feminist perspective to analyze corporate computer vision systems \cite{Buolamwini_Gebru_2018}. As the hype surrounding AI has increased, additional work explicitly linking feminism and AI has emerged. Communications scholar Sophie Toupin \cite{Toupin2023} conducted a critical survey of this literature to create a typology of six ways that feminism and AI have been linked, including feminist ML models, design-based approaches, and feminist influences on AI policy, culture, discourse and science. A recent edited volume, \textit{Feminist AI} \cite{Browne_Cave_Drage_McInerney_2024}, assembled twenty-one chapters ranging in focus from the dearth of women in AI research \cite{Wajcman_Young_2023} to Afrofeminist digital futures \cite{Neema_Chenai_Garnett_2023} to the intersection of AI and racial capitalism \cite{Hampton_2023}. In addition, global feminist networks like the Feminist Internet Research Network \cite{Feminist_Internet} and the <A+> Alliance \cite{Alliance} have emerged to support action on issues of algorithmic bias, labor and the economy, AI-induced gender violence, and more. Civil society organizations have also contributed to the conversation about feminist, decolonial and emancipatory approaches to AI. These are too numerous to comprehensively list, but some examples include Coding Rights in Brazil \cite{Hacking}, IT For Change in India \cite{IT_for_Change}, Data Género in Argentina \cite{Data_Genero_2023}, and Pollicy in Uganda \cite{Pollicy}. Taken together, these show the wide-ranging relevance and utility of feminism for AI research. 

\section{Data Feminism for AI}

In the sections that follow, we review the seven principles of data feminism and explain how they can be adapted to AI research. We also discuss the possibility of two additional principles that address new considerations brought about by AI’s increasing scope and impact on both people and the planet.    

\subsection{Principle 1: Examine Power}

\textit{Data feminism begins by analyzing how power operates in the world.}
\\[5pt]
The first principle of data feminism is to examine power: “the current configuration of structural privilege and structural oppression in which some groups experience unearned advantages—because systems have been designed by people like them and work for people them—and other groups experience systematic disadvantages—because those same systems were not designed by them or with people like them in mind” \cite{D’Ignazio_Klein_2020}. When connecting this understanding of power to data science, we focused on issues of unequal power with respect to minoritized groups--and in particular, on the effects of the under-representation of women and other minoritized groups 1) in the field of data science; 2) as the shapers of research questions; and 3) as the subjects of data-scientific research. The predominance of cisgender men – and the exclusion, even banishment, of women (cis and trans) and nonbinary people, as well as Black, Indigenous, and other people of color, especially when they speak out about AI harms – like Timnit Gebru and Margaret Mitchell – is even more acute in AI research and systems development \cite{Turner_Wood_D’Ignazio_2021}. While \textit{Data Feminism} touched on the role of corporate interest in determining the focus of data creation efforts and research agendas, we were not yet required to contend with the near-total “capture” of AI research and deployment by corporations that has since taken place \cite{Whittaker_2021}. Given this, it is clear that examining power in AI must also centrally involve examining economic power, and the capitalist systems that facilitate the extraction, aggregation, and consolidation of financial resources. 

The capitalism at work in the US, the current epicenter of corporate AI research, has been variously named \textit{surveillance capitalism} \cite{Zuboff_2019}, \textit{oligarchic capitalism} \cite{Foweraker_2021}, and even \textit{neofeudalism} \cite{Dean_2021}. Looking back, we can also see its power emerge in the \textit{racial capitalism} described by Cedric Robinson and others. This is the idea that racialized exploitation has gone hand in hand with capitalist accumulation-indeed that markets are reliant on the production of social hierarchies whereby some groups may own, accumulate, and thrive, and others are excluded, exploited, and marked for premature death. Racial capitalism is also always \textit{gendered capitalism}, as work by Angela Davis \cite{Davis_2011}, Silvia Federici \cite{Federici_2022}, Verónica Gago \cite{Gago_2017} and other Marxist feminists have shown. These are systems in which wage gaps, property laws, gender norms, debt structures, and the lack of reproductive rights conspire to maintain women and genderqueer people as a global economic underclass, with additional repercussions for those who are also Black and brown. In short, capitalism is premised upon the preservation of unequal power: of the enforcement of the racial, gendered, and other social hierarchies which enable the extraction of labor, and therefore value, from the many for the profit of the few \cite{Couldry_Mejias_2019}. These dynamics are clearly visible in the current landscape of AI, in which research agendas are similarly set by the few. It is no surprise, then, that their goal is to maximize corporate profit and to preserve (or even increase) the social and political power that enables it. 

Recent scholarship has also drawn attention to the colonial dynamics of AI research \cite{Couldry_2019, Tacheva_Ramasubramanian_2023, Ricaurte2019}. We see this clearly in the outsourcing of low-paid, traumatizing data-labeling jobs to economically vulnerable people in the Global South which we discuss further in Principle 7 on labor. Here, we emphasize how these capitalist and colonialist dynamics leave so-called "externalities": effects not accounted for in corporate profit models. These include stark environmental impacts, the erosion of job quality, the increased surveillance of workers, a range of harmful incursions on civil rights, outright discrimination, and even death, as we are witnessing right now in Gaza (and discussed in Principle 6, on context). From a macro perspective, when monetary gain is the primary driver of research, we remain in a world in which issues of war, colonialism, racism, and sexism remain unaddressed, since these "externalities" do not contribute to the corporate bottom line. 

The artist Mimi \d{O}n\d{u}\d{o}ha's Library of Missing Datasets \cite{mimimimimi_2024} underscores this point. For \d{O}n\d{u}\d{o}ha, missing datasets serve as “cultural and colloquial hints of what is deemed important” and what is not. When corporations operate without restrictions or regulations, they determine which issues (or groups of people) are worth collecting data about, and which issues (or people) can be ignored. Shareholder interest and public interest are rarely aligned. Corporate choices determine what research questions can be asked, what analysis can be undertaken, which models can be trained, and which users will ultimately be served by those models. We have seen this play out very visibly with respect to LLMs, where researchers such as Emily Bender, Timnit Gebru, Angelina McMillan-Major, and Margaret Mitchell were quick to point out how the models' training data was far from representative, even if its size was substantial \cite{Bender_Gebru_McMillan-Major_Shmitchell_2021}. Subsequent research has documented more specific biases against women \cite{Kotek_Dockum_Sun_2023}, Muslim people \cite{Abid_Farooqi_Zou_2021}, racial and ethnic minorities \cite{Omiye_Lester_Spichak_Rotemberg_Daneshjou_2023, Salinas_Shah_Huang_McCormack_Morstatter_2023}, specific social roles \cite{Lucy_Gururangan_Soldaini_Strubell_Bamman_Klein_Dodge_2024}, and more. 

% While there has been welcome movement towards models trained on more well-documented and intentionally curated datasets (e.g. BLOOM \cite{Workshop_Scao_2023} and MacBERT \cite{Manjavacas_Fonteyn_2022}, the largest language models continue to rely upon undocumented internet data, “correcting for bias” only after the fact. <-- JUST NOTING THAT THIS APPEARS LATER TOO. 

Compounding the issue is how these models continue to be both framed and used as “foundation” models, implying that they serve as a stable basis on which trusted research can be built. Until this misconception is corrected, and likely in perpetuity, we must insist on preventative evaluation, accompanied by an intersectional analysis of power, as a basic starting point. This is in line with the US Office of Science and Technology Policy’s Blueprint for an AI Bill of Rights \cite{Blueprint} and the more recent White House Executive Order on AI, which calls for "robust, reliable, repeatable, and standardized evaluations of AI systems, as well as policies, institutions, and, as appropriate, other mechanisms to test, understand, and mitigate risks from these systems \textit{before they are put to use}" (emphasis ours) \cite{House_2023}. 

These evaluations are necessary for both generative and predictive AI. In terms of predictive AI, in the housing sector, for example, we have seen how automated tenant screening “services” rely on eviction records, arrest records, and credit scores in order to rate tenants on their predicted ability to pay rent. These records are low-quality due to the non-standardized ways in which they are published and scraped, as well as the highly racialized nature of the housing sector in the US \cite{So_2023, Desmond_2016, Nellis_2021, Missing_Credit_2019}. Relying on them disproportionately impacts people of color, and systematically furthers rather than mitigates structural oppressions. 

Here is also where feminist and anti-capitalist critiques of power converge, pointing to the social and historical causes for why certain data sources may be biased against certain populations, as well as to the economic causes for why oppressive and extractive systems exist in the first place. Put another way: a feminist approach to examining power in AI must “ask the other question,” as critical race theorist Mari Matsuda explains, “look[ing] for both the obvious and nonobvious relationships of domination,” and allowing us to “see that no form of subordination ever stands alone” \cite{Matsuda_1991}. In the case of the housing sector, it is both power and profit that drives the informatic asymmetry between the landlords, who can know almost everything about their tenants because they have so much more data and AI on their side, and the tenants, who can know very, very little about their landlords. Tenants do not have the data or the tools to explore, for example, their landlord’s history of evicting people unfairly, or of not maintaining their properties. In short, they do not have the tools to help them build power. A feminist, anti-capitalist approach to AI would focus on designing systems in the service of tenant power – and in the service of all of those excluded by the current rich-get-richer scheme of corporate AI research and development. 
	
\subsection{Principle 2: Challenge Power}

\textit{Data feminism commits to challenging power and working towards justice.}
\\[5pt]
The second principle of data feminism is to challenge the unequal distributions of power that we encounter in the world. 
%In other words: while it is important to understand how power operates in the world today, understanding alone is not enough (nor is describing, measuring, auditing, researching, or explaining). We must also commit to rebalancing that power. 
In \textit{Data Feminism}, we proposed several methods for challenging power in datasets and data projects, as well as for using data science to directly confront corporations and governments. These included collecting counterdata, analyzing data of all kinds with a justice-oriented lens, imagining alternative ways of doing data science, and teaching – laying the groundwork for others to continue this work through data and statistical literacy efforts. In the context of AI, many of these approaches hold. We still need to respond to political demands to collect missing data about underrepresented people and understudied issues. We also need additional ways of devising models and interpreting their output in ways that work towards justice. We need to provide holistic AI education that integrates social and ethical concerns with a healthy dose of history–and crucially, taught by historians and other humanities scholars–so that we stop graduating CS students who are overconfident and underprepared for the complexities of the real world. We need sound laws and policies to reign in corporate power. Finally, we need more creative and participatory and democratic ways to imagine alternative uses of AI, and the space to reject the use of AI if desired or required.  

Examples of each of these aspects to challenging the power of AI are already underway. For example, in New Zealand, a group of Indigenous-led researchers are training an ML-backed speech-to-text system to assist in revitalizing Te Reo Maori, the language spoken by the Maori people \cite{Papa_Reo}. Meanwhile, LLM researchers are working to create explicitly multilingual datasets \cite{CloverSearch} and models for low-resource languages \cite{Lesan}. Of course, more data collection is not always the "solution" to problems of inequality. In many cases, additional data collection can lead to demonstrable harm. This is the “paradox of exposure” that we name in \textit{Data Feminism}, "the double bind that places those who stand to significantly gain from being counted in the most danger from that same counting (or classifying) act" (p 105). So as we celebrate these specific examples, we must also remind ourselves to ask \textit{before} beginning any new project whether a technical intervention is even appropriate, as well as whether we along with the frontline communities we serve have together considered the range of possible harms 
%. In this regard, we can uplift work on data refusal as well as the Feminist Data Manifest-No, a collective effort that elaborates thirty two refusals and commitments related to generating new data futures 
\cite{Cifor_Garcia_Cowan_Rault_Sutherland_Chan_Rode_Hoffmann_Salehi_Nakamura_2019, Zong_Matias_2023}. 

We also need to continue to envision projects that employ AI in the service of justice, such as generating alt-text for images to enhance visual accessibility when it is missing \cite{Alt_Text}. At the moment, however, there are far too few examples of justice-oriented AI work. As has been observed, this may be due to the fact that the data requirements for LLMs and other generative AI models are so large that they cannot account for small-data approaches \cite{klein_2022}. It might also be due to the fact that these models must be trained on data from the past, and the past is rife with structural inequalities that the models learn \cite{Chun_2021}. Furthermore, these models intentionally generate output from the center of any probability distribution, rather amplify lower-probability possibilities. By contrast, feminists contend that outliers-in language as in life-tell us far more than data points at the center \cite{D’Ignazio_2021} \cite{Welles_2014}. This combination of "features" means that both predictive and generative AI are \textit{status quo machines} – truly excellent at reproducing existing conditions and shaping the future in the image of the racist, sexist, ableist, transphobic past. 

Viewed another way, however, the probabilistic basis of these models can be harnessed to challenge and rebalance power. As Wendy Kui Hyong Chun helpfully articulates via the example of models of climate change, these models “offer us the most probable future, given past and current actions, not so that we will accept their predictions as inevitable, but rather \textit{so we will use them to help change the future}” (emphasis ours) (26). Following Chun, what if we treated the biased output of LLMs not as any ground truth but as but as motivation for intervention so that those biases are no longer experienced in the world? 

It is also possible to use both generative and predictive AI to speculate in the Afrofuturist sense–to assist in envisioning otherworldly \cite{Hackett_2023, OBAI} or alternate futures. For example, work led by Wonyoung So \cite{So_2023} employs \textit{reparative algorithms} in order to evaluate which possible interventions might be most effective to close the wealth gap between Black and white families in the United States, an example of how AI can work towards rectifying an unjust status quo.
%His model demonstrates how race-conscious Special Purpose Credit Programs (SPCPs), if deployed as a federal program, could leverage access to housing and credit to support the project of Black reparations. 
%Crucially, the model is not used to determine who should get a loan (the naive, status-quo usage of AI which has been demonstrated to lead to the reproduction of racialized access to mortgages and housing \cite{Martinez_Kirchner_2021}). Rather, modeling is being used to evaluate potential interventions that would best address and 

%: the deeply unequal racialized distribution of wealth in the US. 

At the same time, the scope and scale of the corporate capture of AI also requires a commitment to collective visioning and collective action. There is valuable work being done on data trusts as an alternative to standard data repositories, such as the Worker Info Exchange, which pools data from rideshare workers so that they can ask questions about their employers \cite{Worker}. In the LLM space, we might look to the BLOOM model, an attempt at training a fully-documented large language model in an open, collaborative way \cite{Workshop_Scao_2023}. Researchers are also considering how they might form independent research coalitions in order to lobby for access to the data of Big Tech that they need in order to conduct their work \cite{Home}. And workers whose livelihoods are being threatened by AI are turning to unions and organizing in order to advocate for the conditions they need in order to thrive \cite{How_Hollywood, Hollywood_Actors}. 

These interventions at the level of civil society must be accompanied by government regulation and policy. The EU AI Act, if implemented with force, promises to become a powerful tool to limit the incursion of Big Tech into personal lives, as does the US Executive Order on AI–for as long as it remains in effect. Some countries, such as Canada \cite{Canada_2021} and Denmark \cite{In_Search}, have gone even further, incorporating explicitly feminist and anti-oppressive policies into their governments. We will, of course, always continue to require resistance in the form of care, community, solidarity, and mutual aid – nothing less than a recognition of the shared humanity that capitalism would have us forget. If we are going to resist the complete capture of AI by capitalist forces, then we must return to these emphatically human and anticapitalist models as our guides. 

\subsection{Principle 3: Rethink Binaries and Hierarchies}

\textit{Data feminism requires us to challenge the gender binary, along with other systems of counting and classification that perpetuate oppression. }
\\[5pt]
The third principle of data feminism derives from the false binary that Western culture has constructed between the category of “man” and the category of “woman.” There are more than two genders, of course, and a fundamental commitment of feminist thought is to equality for all genders. Furthermore, binaries are often hiding hierarchies, and the gender binary is a key example. It hides a hierarchy-patriarchy-in which cisgender men are on top, dominating social institutions from corporate boards to government leadership positions; women, trans, and genderqueer people are intentionally kept on the bottom; and there is no space at all for anyone in between. One need only look at the skewed gender balance of the signers of the “AI pause” letter, those invited to testify before the US Congress about AI, or the placement of Larry Summers–infamous for his derogatory comments about women and science \cite{Summers}–at the head of the reconstituted board of OpenAI, in order to see the obvious issues with inclusion in AI. But the issues with the gender binary and with the hierarchy it maintains run much deeper–into the formation of the field of AI research.

%, the way in which its own history is narrated, and into the naive computational approaches to gender–and to the classification and measurement of social categories more broadly–that dominate the field.

The field of data science–the focus of \textit{Data Feminism}–has never been known for its exemplary inclusion. But the scope and scale of the conversation around the power and perils of data did lead to an observable change–if not a full recalibration–in inclusivity in the field. 
%In recent years, an array of initiatives have emerged around topics of data literacy and computer science education, as well as data science for the public good. 
In under a decade, Harvey Mudd College, under the leadership of president Maria Klawe, increased the percentage of women computer science majors to over 50\%, demonstrating that CS professors and administrators that whine about "the pipeline" have no leg to stand on \cite{Harvey, Klawe_2013}. 
%The University of Virginia’s new School of Data Science, founded in 2019, is explicit about its view of “justice as a key concept in this endeavor.” 
But it is not a coincidence that as the broader fields of data science and computer science have grown more inclusive, a subset of researchers from within those fields have begun to pull away in order to reconstitute the field of AI. That this group is WEIRD–Western, educated, industrialized, rich, and (arguably) democratic \cite{Septiandri_Constantinides_Tahaei_Quercia_2023}–and also heavily dominated by white cisgender men, appears to us as evidence of another iteration of a common trend in technical fields: the emergence of new gatekeeping mechanisms in order ensure that those at the top are able to maintain their elite status in the field. 
%In Data Feminism, the authors point to the early history of computing and the emergence of computer science degrees to credential men in technical work–and displace the women secretaries who had previously worked as human “computers”--and draw parallels to the field of medicine in which men with medical degrees and wacky theories replaced women midwives whose work was grounded in empiricism \cite{Witches}. Here is this same process taking place, once again, before our eyes.  

We also see this power move taking place on a conceptual level, in which consumers of AI technologies are being encouraged to believe that its capabilities are magical and mysterious, and thus impossible for any normal person to understand. 
%It is true that many AI technologies are indeed complex–one need only point to the 950 coauthors of the paper announcing Gemini, Google’s multimodal LLM–in order to affirm that the development of these models requires a range of expertise. But maintaining the mystery of such models also serves a strategic purpose: it makes it appear to end-users that they should not question or make demands of such tools; to researchers that they should not demand any documentation or ability to peek under the hood; and to government regulators that they lack the knowledge to draft policy or legislation to govern AI. 
This, too, has a historical antecedent in the way that software developers became figured as “wizards” and “sorcerers” in a way that made the acquisition of technical expertise seem off-putting to those without access to formal training mechanisms \cite{Ensmenger_2012, Hicks_2018}.
%. It also served to exclude those who actually did the work of founding the field \cite{Programmed_Inequality}. 
But it also serves a strategic purpose: it makes it appear to end-users that they should not question or make demands of such tools; to outside researchers that they should not demand any documentation or ability to peek under the hood; and to government regulators that they lack the knowledge to draft policy or legislation to govern AI. We run the risk of repeating this pattern when we accept the language of AI’s “emergent properties” and the hagiography of the field’s supposed founders, as echoed in a 2023 New York Times article that elevated exclusively men, mostly ultra-rich and white \cite{Kim_2023}. 
%It is evident even in the stories we tell ourselves about the field, as Keyes and Creel’s work on the feminist philosopher Alison Adam helps show. 
This is why the inclusion implied by the slogan of the Algorithmic Justice League is so important at this present moment: "If you have a face, you have a place in the conversation." \cite{Algorithmic_Justice_League}.  
%Otherwise said, AI is touching all of our lives and thus all of us should be at the table, not just the white posterboys of wizardly innovation. 

It is also important to recognize the very real ways the gender binary has been operationalized and even weaponized in AI research. Beyond ample evidence of the gender biases against cisgender women, and trans and nonbinary people, that are entrenched in LLMs, AI researchers appear to have an antiquated understanding of gender itself \cite{Keyes_2018, Devinney_Björklund_Björklund_2022, Stanczak_Augenstein_2021, Scheuerman_Paul_Brubaker_2019}. This lack of understanding of gender, in turn, leads to the inaccurate (at best) and harmful (at worst) research design and applications. And sex fares no better
%. In a 2022 paper, Kendra Albert and Maggie Delano break down three common problems in using sex and gender data in the context of electronic health records, as well as how those problems result in harmful outcomes for trans and nonbinary people navigating the medical system. All of these issues are based on "false assumptions that sex and gender are binary, static, and concordant" 
\cite{Albert_Delano_2022}. There is a near-universal failure to understand that neither sex nor gender are essential, "natural", or fixed properties, and they certainly cannot be "detected." 
%As many theorists have pointed out (and it appears most AI researchers have failed to read them) gender (and race) are social technologies in their own right \cite{Benjamin_2016, Chun_2009, Coleman_2009, Sheth_2004}. 

One might extend this gender trouble to other projects that seek to identify, measure, or otherwise quantify complex aspects of social and physical difference. From the entire industry being built around “Emotion AI,” which seeks to measure emotion in images and in real-time video, to employment screening services that purport to measure employability, to what has been termed \textit{neo-phrenology}: an attempt to measure emphatically immeasurable qualities such as intelligence or criminality using algorithmic means \cite{Dark_Past_2020}. 
%In Lucknow, India, police proposed to use computer vision to monitor women's facial expressions in public spaces for signs of distress \cite{Chandran_2021}. Critiquing such forms of paternalistic intrusion (and fortification of carceral infrastructure), STS scholar Radhika Radhakrishnan wrote, "surveillance not only fails to keep women safe, it produces its own kind of violence" \cite{Radhakrishnan_2021}. 
In these and any new proposed use cases, we must ask ourselves whether they are reinforcing binaries – gender binaries, patriarchal hierarchies, and paternalistic stereotypes – and if so, instead think otherwise about how to model gender, how to build tools for gender safety, and how to create a world where technologies cease to erase and harm trans and non-binary people. 

\subsection{Principle 4: Elevate Emotion and Embodiment}

\textit{Data feminism teaches us to value multiple forms of knowledge, including the knowledge that comes from people as living, feeling bodies in the world.}
\\[5pt]
Elevating emotion and embodiment with respect to data is connected to the previous principle, because emotion is often viewed in binary opposition to reason, as a "feminized" element that should be exiled from scientific research because it is subjective as opposed to objective; "soft" (feminine stereotype) instead of "hard" (masculine stereotype). With a critique of hidden hierarchies in mind we can also return to the binary between reason and emotion and recognize 1) how it is a false binary, and there is no such thing as purely rational science \cite{Keller_1996}; and 2) how this binary is hiding a hierarchy, in which emotion, embodiment, lived experience, and other feminized ways of knowing are relegated to lower status forms of knowledge. 

The tools to challenge this false binary come from feminism. Feminist philosopher Donna Haraway’s concept of \textit{situated knowledges}–the idea that knowledge originates at a particular time, in a particular place, and from within a particular set of social and political contexts–helps us recognize how all knowledge, including scientific knowledge, is shaped by the perspectives of the people who produce it \cite{Haraway_1988}. More recently, Black feminist theorist Katherine McKittrick points us to how there are multiple systems of knowledge-making, and to how these must be understood as sites of "collaborative praxis" in and of themselves, and as relational with respect to one another \cite{McKittrick_2021}. Crucially, a refined awareness of multiple knowledge systems enhances, rather than detracts from, our collective knowledge. 

This \textit{feminist objectivity} bears relevance to AI research in myriad ways, not the least in how the conversation around the risks and harms related to AI has unfolded over the past year or so. On the one hand, we have seen those in positions of power (who, not coincidentally, themselves represent dominant social groups) sound the alarm on future hypothetical harm–of AI robots developing nuclear weapons and other questions of moral “alignment.” On the other, we have seen AI researchers with equal technical expertise, enhanced by the “empiricism of lived experience” as women (cis and trans), queer people, and/or people of color, calling attention to the harms being perpetuated by AI systems right now \cite{Conference_Programme_2018}. Famously, Dr. Joy Buolamwini, a Black woman computer scientist, undertook an "evocative audit," and used the example of her own face to demonstrate the poor accuracy of facial recognition software. She then connected that evidence to the greater harms being unleashed as these inaccurate software systems were (and remain) deployed in police departments which are already inextricable from structural racism \cite{Buolamwini_2023}. 
%By the same token, the FTC recently barred RiteAid, the national drug store chain, from using facial recognition technology as a method of detecting shoplifting after an audit determined that it discriminated against Black and Latinx customers \cite{Rite_Aid_2023}. 
We have also seen research documenting the misogyny and racism, as well as instances of rape and pornography, in multimodal datasets such as LAION, resulting in LAION pulling its datasets from circulation \cite{Goldman_2023}. Very crucially, this research was first performed by Black women-led research teams who brought their personal experience with these datasets to their research. Yet it was only when this research was replicated by white researchers at Stanford that the team behind LAION took note. This is unfortunately already a pattern in AI research in which Black women who bring their embodied experience to their research have their concerns dismissed, only to have those concerns validated by people from dominant groups, often years later \cite{Bender_Gebru_McMillan-Major_Shmitchell_2021, Turner_Wood_D’Ignazio_2021}. 

It is not just that the full range of human experience should be incorporated into assessments of AI harms. Another binary that must be challenged has to do with our perception that work with data and AI can only serve to increase efficiency, automation, and control. 
%While classical sociological accounts of data have theorized that it is a “technology of a distance” \cite{Porter_2020} or a tool for taming territory \cite{Scott_1998}, more 
Recent research describing the data practices of feminicide data activists has shown how the production of data can become a tool of intimacy, relationality, care, and memory work \cite{Gatwiri_Hasanova_2023}. Scholar and activist Helena Súarez Val has described these works as "affect amplifiers," translating feminist grief and rage into public action \cite{Val_2021}. 
%Similarly, the project team behind Homegoinng, an online memorial to the victims of Covid-19, describe their experience of reattaching lives to numbers in a spreadsheet as feminist care work, and a crucial component of data analysis \cite{Homegoing}.
This form of collective, community-engaged, intentionally emotion-laden work can also remind us of our 
%Valuing emotion and other forms of lived experience is not only about individual bodies, but also about 
interdependence, a key idea from disability studies that emphasizes how
%which "reminds us that we rely on each other, and our actions can have consequences on others" 
"we rely on each other, and our actions can have consequences on others" \cite{Ansari}. 
%We must continue to resist the exclusion of people with diverse needs through intentionally grounded and at times creative work. This may mean leveraging existing systems that already work, as Meredith Broussard has documented in her interview with Richard Dahan, a Deaf Apple Store employee whose request for live (human) captioning was rejected on the grounds that AI could do it more cheaply–but far less effectively, as it turns out \cite{Broussard_2023}. It may also mean thinking creatively and expansively about how the unique capabilities of generative AI might be enlisted in the service of  a wider range of bodies and minds \cite{Whang_2023}. But it may also mean 
As we move forward, we must continue to consider how to craft AI systems that focus on "practicing alliance," create "carewebs and pods," and are designed for "misfitting.” \cite{We_Are_Here, Piepzna-Samarasinha_2018, Garland-Thomson_2011}. Emerging steps towards such systems are happening in human computer interaction with conversations around trauma-informed computing, healing databases, and restorative/transformative data science, and should be expanded to AI research. 

\subsection{Principle 5: Embrace Pluralism}

\textit{Data feminism insists that the most complete knowledge comes from synthesizing multiple perspectives, with priority given to local, Indigenous, and experiential ways of knowing. 
}
\\[5pt]
The principle of embracing pluralism builds from the previous principle, which emphasizes alternate forms of knowing. Here, the emphasis is on how we might represent multiple perspectives in our work. Crucially, “multiple perspectives” does not simply mean multiple opinions. Building on the work of Sandra Harding, and more recently, the Design Justice Network, “multiple perspectives” has become a beacon for the wide range of experiences, social positions, and places in the world from which people produce knowledge. The central premise of embracing pluralism is that we can gain better, more detailed, more accurate, and ultimately more truthful knowledge if we pool these perspectives together, paying particular attention to the perspectives of those who are most directly impacted by the issue at hand.

Embracing pluralism is of crucial importance in the context of AI research, both because of the narrow demographic composition of AI researchers themselves (see Principle 3, on binaries and hierarchies), and because of the imbalance of power between those currently designing AI systems, and those subject to their decisions (see Principles 1 and 2, on power). Put simply: the field of AI needs to develop more participatory, more responsible, and more humble methods for being in dialogue with impacted communities. 
%While other fields have long traditions of community engagement (e.g. urban planning, digital humanities, design), AI remains anchored in values that are antithetical to sustained collaboration, e.g. novelty, efficiency, and generalizability \cite{Birhane_Kalluri_Card_Agnew_Dotan_Bao_2022}. 
Catherine had the opportunity to learn this first-hand in a project involving the co-design of AI tools in collaboration with grassroots feminist data activists across the Americas. Together, they decided to develop an ML classifier for detecting news articles about feminicide, a systematic violation of human rights that involves the gender-related killings of women and girls. The tool was intended to streamline the work of the activists. 
%undertaken by data activists to manually log details about every case of feminicide into spreadsheets and databases.  In the brainstorming phase, the research team proposed automating not only the detection of feminicide news articles but also the extraction of information from news articles into activists' databases. 
But the activists pushed back on fully automating the process for two reasons: (1) the impossibility of generalizing any definition of feminicide, and (2) because of how they saw their work
%. First, they noted that different activists use different definitions of feminicide in different places, making a generalized approach inappropriate. And secondly, activists spoke about their careful, investigative work 
as a form of memory justice: a way to honor the lives of the women killed so that they would not be forgotten. 
%Activists did not want to automate this labor (indeed, they felt that the important witnessing, remembering and care that they took with people's lives could not be automated). 
The result was a tool that 
%was co-designed, that honored the memory work of the activists but still reduced the extraneous violence that they were exposed to, and therefore 
has seen impressive uptake by human rights monitoring groups around the globe. This stands in contrast to the numerous ML/AI tools created without consultation, which lack both users and use.     

%When shifting from smaller-scale ML approaches to LLMs and other current AI technologies that rely on big data and significant compute, 

The prospect of embracing pluralism becomes harder in terms of both architecture and access when scaling up to larger models. LLMs must be pre-trained by those with access to both data and compute, introducing barriers to early-stage participation that had previously not existed. While the push for more open-source models (e.g. BLOOM, OLMo) and more transparent models (e.g. the Stanford Transparency Index) are certainly welcome, the technical requirements of training such models make it far more difficult to involve impacted individuals and small groups in their creation. 
% We will  continue to see important developments in terms of decreasing model size and more efficient methods of fine-tuning, but these solutions to broadening access remain focused on technical (if non-corporate) users. We also do not yet have sufficient examples of involving impacted groups in these models’ design and deployment–or, for that matter, examples of their utility in those contexts. 
One recent example that suggests a path forward comes from a project led by Maria Antoniak \cite{Antoniak_Naik_Alvarado_Wang_Chen_2023} which surveyed both pregnant people and healthcare providers about their ideas and fears about the possibility for an AI-assisted chatbot that might help to navigate the experience of pregnancy. Like the feminicide classifier, Antoniak et al. very crucially did not build the chatbot first and then ask for feedback; instead, they created a study that included people with multiple forms of expertise and that enabled them to share that knowledge. The output of the project was a set of values, distilled from the participants themselves, that might guide future work. 

There is another level at which we might conceive of pluralism, which is at the scale of government and those democratic bodies which have been designed to serve the public interest. We see this approach to embracing pluralism in the ideas of public-interest AI \cite{Broussard_2023} and digital public infrastructure \cite{What_Is_2020}. Like physical infrastructure – roads, parks, schools – these systems would be conceived and designed through public processes, guided by the public interest, and come with transparency and governance requirements. While we have seen important lawsuits and other regulatory processes directed at corporations and the technologies they develop, we might also envision citizen-led design processes in which ideas are sourced from communities and projects are funded (and maintained) with taxpayer dollars.

%Returning to the example of tenant-screening services, how might we design a system that screens landlords for their history of property maintenance or illegal evictions (similar to the Anti-Eviction Mapping Project's Evictorbook app but at the scale of the country Anti-Eviction Mapping Project)? While the modeling approaches exist to create such a system, and it might be possible to harness public records to create sufficient data, corporate incentives would never lead to such a system being built. 

These examples of participatory design notwithstanding, it is important to acknowledge that “participation is not a design fix for machine learning,” as Mona Sloane et al. have explained \cite{Sloane_Moss_Awomolo_Forlano_2020}. Participation can be tokenizing, extractive, under-resourced by researchers inexperienced with the investment required, or incorporated too late in the design process to influence its outcomes, among other issues that the authors raise. We must continue to push back against what Lorraine Daston might call the “epistemic virtues” of AI/ML research, which echo those of the larger field of computer science: its emphasis on novelty, generalizability, and efficiency, which not only over-determine the kinds of research that are support but are also implicitly believed to be best \cite{daston_2014}. 
%A final note about embracing pluralism relates to the field of computer science itself. In addition to opportunities to involve the general public, it is also important to recognize how research itself can benefit from involving multiple disciplinary perspectives. 
While much attention has been devoted to the capture of technical research by corporations, it is also true that there exists a capture of academic research by the field of computer science. Those of us in academia have observed the trends in the composition of college majors, and resultant faculty hiring, leading to outsized resources being invested in computer science and the forms of research it supports to the detriment of all other fields. 

Computer science continues to push out women and people of color at a scale not seen in other sciences. It systematically fails to teach the value of context-sensitive, culturally appropriate technology development, and engages in masculinist fantasies of foundational models ("one model to rule them all"). The singular hubris of this field seems to know no bounds, even as evidence of harm surfaces over and over again. We have also observed how fields such as HCI and social computing draw from expertise in, for example, anthropology, sociology, and literary studies, even as CS departments refuse to hire scholars with PhDs in those fields to train their students. 
%Meanwhile, in the LLM space, scholars in the humanities are brought in as domain experts but not valued for their foundational theories of language. 
If we are sincere in our belief that the best, most accurate, most truthful knowledge comes from the pooling of multiple perspectives, then we must restructure our own research processes and funding models in order to account for this fact.    

\subsection{Principle 6: Consider Context}

\textit{Data feminism asserts that data are not neutral or objective. They are the products of unequal social relations, and this context is essential for conducting accurate, ethical analysis. }
\\[5pt]
The sixth principle, to consider context, is at once universally applicable to AI research and yet (nearly) universally ignored. This principle applies most directly to the issue of training data, which we began to  discuss with respect to language models in the principle on power. Related work has shown how additional (human) decisions made during the training and filtering process introduce additional biases into these datasets, such as for text written by those of higher socioeconomic status \cite{gururangan-etal-2022-whose}, and those who occupy specific social and professional roles \cite{Lucy_Gururangan_Soldaini_Strubell_Bamman_Klein_Dodge_2024}. Related research has explored text-to-image models, with findings related to the reproduction of sexist, racist, and colonial biases \cite{Nicoletti_Equality_2023, Qadri_Shelby_Bennett_Denton_2023}. What we learn from feminism is how understanding the contexts in which these datasets are created, and more broadly, a recognition of how all data is shaped by unequal social relations, is essential for identifying any downstream biases or potential harms. It is also necessary for ensuring that research questions are properly framed, that evidence is properly analyzed, and that any claims that might be made on the basis of that analysis are properly scoped.     

There exist several valuable recommendations for how to restore context to ML models more broadly, such as Eun Seo Jo and Timnit Gebru’s suggestion to adopt practices similar to those employed by archivists when curating and documenting datasets \cite{Jo_Gebru_2020}, and Mitchell et al.’s proposal of “model cards for model reporting” when releasing a model for general use \cite{Mitchell_Wu_Zaldivar_Barnes_Vasserman_Hutchinson_Spitzer_Raji_Gebru_2019}. We would be well-served by considering how these might be adapted to the current AI/ML landscape, particularly given that neural-network-based architectures of generative AI models make it difficult for the end user (or anyone) to trace any specific model output back to the sources that contributed to it. But to do this work, we must challenge the present stratification of labor in the data science pipeline – the erroneous idea that curating, labeling, and documenting data is somehow unskilled labor and that the analysis and modeling part of the pipeline is where the "science" is at \cite{Feinberg_2022}. 
%As Sambasivan et al state, "Everyone wants to do the model work, not the data work." \cite{Sambasivan_Kapania_Highfill_Akrong_Paritosh_Aroyo_2021} 
As archivists and others in the humanities know well, the work of creating a dataset that accurately represents the research question at hand is long, painstaking, and premised on deep expertise. Yet the dominant mantra of Big Tech remains “move fast and break things.” This fundamental mismatch of value folds back into the capitalist critique that began this paper; so long as the production and curation of data is not seen as valuable, then we will not be able to sufficiently support this type of human expertise. Moreover, the AI systems that we produce with naive and ill-considered data will just plain get things wrong. These are the "data cascades" that Sambivasan et al. identify as amplifying into major quality problems downstream \cite{Sambasivan_Kapania_Highfill_Akrong_Paritosh_Aroyo_2021}. 

%Another aspect of considering context relates to  the broader social context in which algorithmic decisions are made. Work by Ayanna Howard has famously shown the risk of automation bias: the belief that any algorithmic or “automated” decision is more truthful than a human decision. She used an experiment that simulated an emergency evacuation from a building and showed how people would more readily follow a robot than a human, even when the robot led them closer to the danger \cite{Robinette_Li_Allen_Howard_Wagner_2016}. Related work by Ben Greene has shown how humans layer in social bias on top of automation bias, for example when choosing to believe algorithmically generated risk scores when they penalized Black defendants, but to ignore them when they penalized whites \cite{Green_Chen_2019}. These are additional contexts that we must account for designing AI systems as well as deploying them, carefully considering how we might convey these models’ limitations along with their strengths. <-- CUTTING THIS WHOLE GRAF FOR NOW, NOT SINCE IT'S IMPORTANT BUT BECAUSE IT SEEMS LIKE NO ONE WILL MISS IT IF IT'S NOT THERE 

A final consideration with respect to context has to do with historical context–and in particular, of the historical context of AI research itself. This history is an uncomfortable one, as it points to the longstanding complicity of the field of computer science with US military research. This dates back to the DARPA-funded creation of ARPANET, the precursor to the internet \cite{darpa}. It carries through to the construction of OntoNotes, the hand-labeled language model underlying many common NLP libraries, which was another DARPA-funded project \cite{4338389}; and to specific approaches to language modeling, such as topic modeling, which came about through a government desire to monitor global newswire messages at scale \cite{binder_2016}. More recently, news reports have drawn our attention to the role that Amazon has played in cloud storage for the US Immigration and Customs Enforcement agency (ICE) \cite{hao_2018}; and how Google developed technology to make drone strikes more accurate, which they then sold to the US government \cite{Project_Maven}. These developments are no longer theoretical, as evidence emerges that they are being put to use by Israel against Palestine, and with horrific human costs \cite{Davies_McKernan_Sabbagh_2023}. As advocacy efforts such as \#NoTechForApartheid \cite{notechforapartheid} and publications such as Logic(s) Magazine \cite{Logics} powerfully remind us, these inhumane, destructive, and–in the context of Palestine, genocidal–contexts are those in which far too many technical innovations are put to use. We, as AI researchers, can no longer claim ignorance about these contexts as among the uses for our work.

\subsection{Principle 7: Make Labor Visible}

\textit{The work of data science, like all work in the world, is the work of many hands. Data feminism makes this labor visible so that it can be recognized and valued. }
\\[5pt]
This final principle highlights the many people whose labor enables work with data. In \textit{Data Feminism}, we observed how the work of data science replicated professional hierarchies, with credentialed data scientists at the top, and those perceived to occupy less technical roles–such as data annotation and content moderation–on the bottom. We also observed how this professional hierarchy could be mapped onto gendered, raced, and ultimately colonial hierarchies, with those in the Global North occupying the high-status and high-compensation roles, and those in the Global South occupying those at the bottom. 

Because the current configuration of AI research is premised upon the consolidation of significant resources–technical and economic as well as human–this colonial structure has become solidified as the fundamental framework on which AI depends. One need only look at the investigative reporting that followed the initial release of ChatGPT, which showed that this “artificial” intelligence depended on very human workers in Kenya screening potentially offensive responses in real-time \cite{Exclusive_2023}. 
%Or the continued coverage of RLHF systems, which consistently trace this “human feedback” to workers in the Global South \cite{Mediation}. 
The location of these workers is not a coincidence. Scholars such as Julian Posada have asserted that companies in the Global North exploit political instability and capitalize on catastrophe to enrich themselves, a familiar and longstanding historical pattern \cite{Posada_2022}. This joins a long line of research (and evidence in the world) that documents how capitalism is fundamentally dependent upon resource extraction–and on paying as little as possible for those resources, including human labor, in order to maximize profit \cite{Ricaurte2019, Couldry_Mejias_2019}. 
%This was the economic motivation that led to the early entrenchment of chattel slavery, and is still visible in the range of forms of racial capitalism that remain in place today.    

The workforce of the Global North is not exempt from the incursion of AI, of course. In the US, we have seen lawsuits by Getty Photography and The New York Times brought against companies peddling generative AI technologies. In addition, the summer and fall of 2023 saw major strikes by the Writers Guild of America \cite{How_Hollywood} and the Screen Actors Guild \cite{Hollywood_Actors}. These culminated in necessary protections against the use of AI to revise scripts and in the creation of likenesses of human actors. We also celebrate the larger trend towards collective action in the tech sector, and in white collar jobs more broadly, as a counterforce to the relentless individualism championed by capitalism. Graduate students at elite engineering universities in the US are beginning to unionize, joining long-standing unions at public institutions including the University of California system, the Cal State system, and the City University of New York, among others. We must continue to ensure that these efforts at building solidarity cut across lines otherwise drawn by technical expertise. After all, the history of colonialism tells us that those at the lower end of labor hierarchies will be the ones most impacted by any move to increase profit or workplace efficiency. Without solidarity across class, race, gender, and work sector, capitalist power will only continue to accrue.  

\section{Discussion: Future Principles for Feminist AI}

The previous pages document our current thinking about how the seven principles defined in \textit{Data Feminism} can be applied to AI research, but two topics require additional attention: the environmental impact of AI research and deployment, and issues surrounding consent. Here we summarize our present thoughts on each.

\subsection{Data Feminism and the Environment}

In many ways, questions about the environmental impact of AI follow from how its development and deployment reinforce historical patterns of capitalism and colonialism. Resource extraction, after all, is as much about natural resources as it is about human labor. It has long been observed that the environmental impacts of this resource extraction are experienced unequally, with people in the Global South experiencing the deleterious effects of climate change in far greater measure than those in the Global North, even as they contribute far less to global emissions. Google, for example, used 5.6 billion gallons of water in 2023, up 20\% from the prior year \cite{Google_Thirsty_2023}. An average Meta data center consumes as much electricity as 150,000 average homes \cite{say_2023}. As current research into the energy and water requirements of LLMs has shown \cite{li2023making}, AI seems positioned to further exacerbate these effects. Again, these systems seem positioned to benefit elite users in the Global North, even as they exact their cost on those in the Global South. This is an environmental issue, but it is also a feminist issue, as these effects are not only experienced unequally in terms of geography, but also in terms of gender. A feminist principle for AI about the environment might draw from the several decades of ecofeminist scholarship which has worked to establish the connection between environmental harms and other forms of structural oppression. It might also look to work by Indigenous feminists in Latin America, who view the “cuerpo-territorio” (body-land) as an interconnected system, and by North American Indigenous feminists who similarly link body and land sovereignty while working towards the end of structural violence against both \cite{Cabnal_2010, Deer_2015, Lucchesi_2022, Simpson_2017}. 

\subsection{Data Feminism and Consent}

Consent is also a longstanding feminist concern due to the high rates of rape and sexual violence faced by women, trans and non-binary people around the world living under cisheteropatriarchy. Most Western laws that address rape and sexual violence have their basis in some form of consent, and there are various feminist formulations of what that might mean, such as the popular 2016 FRIES model from Planned Parenthood where consent is: Freely given, Reversible, Informed, Enthusiastic, and Specific. With that said, there are numerous feminist critiques of consent as being too individually focused, too simplistic, and too binary (e.g. reinforcing heteronormative, gendered stereotypes of aggressive men and gatekeeping women and ignoring queer relations entirely) \cite{Alcoff_2018, Edenfield_2019, MacKinnon_2016}. Because of the violence and harms already being propagated by AI systems, and because of the likelihood that these harms will continue to increase, we still think it may be useful to formulate a feminist principle for AI about consent in expanded forms, i.e. queer, collective and/or interdependent consent. 

AI is currently facilitating the mass, non-consensual exploitation of pornographic images of women in the form of deep fakes. 
%The award-winning 2023 film Another Body recounts the story of a college student who was alerted to pornographic videos posted across the internet with her face deep faked onto another person's body, along with striking statistics about the growing prevalence of this violence and the impunity that surrounds it. 
As Danielle Citron teaches us, this is consistent with a number of "intelligent" technologies such as networked cars, mobile phone apps, and more which have been exploited by abusive perpetrators to control and dominate their partners\cite{Citron2022-lx}. Then, there are the broader issues of consent that relate to the data sources used to train LLMs and generative AI systems, as discussed in Principle 6. Are online data–including social media posts, family photos, original artwork, journalistic reporting and personal blogs–fair game for inclusion in massive data sets without the creators' knowledge? As we await the development of informed guidelines for fair use, we can be certain that something other than the current system–in which Big Tech steals people's work, exploits it, makes money, and facilitates structural violence along the way-is required. There has already been significant work around consent and technology, including the Consentful Tech Project by Allied Media Projects, and work on what consent means in human computer interaction \cite{Im_Dimond_Berton_Lee_Mustelier_Ackerman_Gilbert_2021, Lee_Toliver_2017, Strengers_2021} which we can use to build more relational, inclusive, and liberating systems--and to reject them if they are not. 
%At its core, consent must be an intimate, dynamic process, premised on relationality and non-violence.

\section{Conclusion}

While the forces of racial, gendered capitalism that are currently shaping AI research are powerful, they are also predictable. They operate in ways we have observed and experienced for centuries. We offer these thoughts on feminist principles for AI research, along with our hope, because we know what will happen if we do not change course. The predictability of late-stage capitalism, in some ways, gives us an easy goal: if the status quo is not what we what, then we must follow Ruha Benjamin's call to “craft the worlds you cannot live without, just as you dismantle the ones you cannot live within” \cite{Benjamin_2022}. 

\begin{acks}
We would like to thank Nikki Stevens and Isadora Cruxên for their feedback on early drafts of this paper. This work has been partially supported by a grant from the Mellon Foundation (G-2211-14240).
\end{acks}

\bibliographystyle{ACM-Reference-Format}
\bibliography{df-for-ai-bib}

\appendix
\section{Research Ethics and Social Impact}
\subsection{Ethical Considerations Statement}
This paper is primarily a literature review and theory contribution, thus we did not engage in any human subjects research, systems development or deployment. Our work has been guided by considerations for citational justice, and we have specifically sought to cite the work of scholars from marginalized backgrounds, especially BIWOC and queer people of color, as well as forms of knowledge production not recognized by the academy, including activism, journalism, art, design and creative communication projects. 

\subsection{Researcher Positionality Statement}
We are a two-person writing team that brings domain expertise in the humanities (including historical and literary scholarship), digital humanities (including NLP/ML), urban planning, software development, data science, and data visualization. As two cisgender women, we share an interest in and commitment to gender equality for all genders which draws us to the intersectional, transinclusive feminist frameworks outlined in this paper. That said, there are many experiences of intersectional oppression that we do not have – we are white settlers, cisgender, (mostly) heterosexual, (mostly) non-disabled. For these reasons, we have tried to listen to, learn from and cite authors who write from these intersections. We strongly believe in the logic of co-liberation. As the Combahee River Collective states, "None of us are free until all of us are free." 

\subsection{Adverse Impact Statement}
With this work, we seek to spark conversations across disciplines about the social and political inequalities being exacerbated by AI. Rather than "calling out" specific people or institutions, we hope this critique serves as a "call in" other scholars to join together and work collectively towards building the AI systems that we all deserve.

\end{document}